\documentclass[fleqn,usenatbib]{mnras}

\usepackage{newtxtext,newtxmath}
\usepackage[T1]{fontenc}

\DeclareRobustCommand{\VAN}[3]{#2}
\let\VANthebibliography\thebibliography
\def\thebibliography{\DeclareRobustCommand{\VAN}[3]{##3}\VANthebibliography}

\usepackage{graphicx}   % Including figure files
\usepackage{amsmath}    % Advanced maths commands
\usepackage{threeparttable}
\usepackage{subfigure}
\usepackage{multirow}
\usepackage{multicol}
\usepackage{booktabs}
\usepackage{makecell}
\newcommand{\swi}{{\it Swift}}
\newcommand{\fer}{{\it Fermi}}
\newcommand{\maxi}{{\it MAXI}}

\title[Torque reversals and wind variations of X-ray pulsar
Vela X-1]{Torque reversals and wind variations of X-ray pulsar Vela X-1}

\author[Z. Liao et al.]{
Zhenxuan Liao,$^{1,2}$
Jiren Liu,$^{3}$\thanks{Email: liujiren@bjp.org.cn}
Lijun Gou$^{2,4}$
\\
$^{1}$School of Information Engineering, Sanming University, Jingdong Road 25, Sanming 365004, Fujian Province, People's Republic of China\\
$^{2}$Key Laboratory for Computational Astrophysics, National Astronomical Observatory, Chinese Academy of Sciences, \\~~Datun Road 20A, Beijing 100012, People's Republic of China\\
$^{3}$Beijing Planetarium, Xizhimenwai Road, Beijing 100044, China\\
$^{4}$School of Astronomy and Space Science, University of Chinese Academy of Sciences, Beijing 100049, People's Republic of China\\
}

\begin{document}
\label{firstpage}
\pagerange{\pageref{firstpage}--\pageref{lastpage}}
\maketitle

\begin{abstract}
The erratic spin history of Vela X-1 shows some continuous 
spin-up/spin-down trend over tens of days.
We study the orbital profile and spectral property of Vela X-1
in these spin-up/spin-down intervals, using the spin history monitored
by \fer/GBM and light curve from \swi/BAT and \maxi/GSC. 
The BAT fluxes in the spin-up intervals are about 1.6 times those of the 
spin-down intervals for out-of-eclipse orbital phases.
The spin-up intervals also show a higher column density than the spin-down intervals, 
indicating there are more material on the orbital scale for the spin-up intervals.
It could be due to the variation of the stellar wind of the optical 
star (HD 77581) on tens of days. The varying wind could lead to 
alternating prograde/retrograde accreting flow to the neutron star, which 
dominates the transfer of the angular momentum to 
Vela X-1, but not the total observed luminosity. 

\end{abstract}


%%%%%%%%%%%%%%%%%%%%%%%%%%%%%%%%%%%%%%%%%%%%%%%%%%

%%%%%%%%%%%%%%%%% BODY OF PAPER %%%%%%%%%%%%%%%%%%

\section{Introduction}
\label{sec:intro}

%-------------------------------------------------------------
\begin{figure*}
    \centering
    \includegraphics[width=0.92\textwidth]{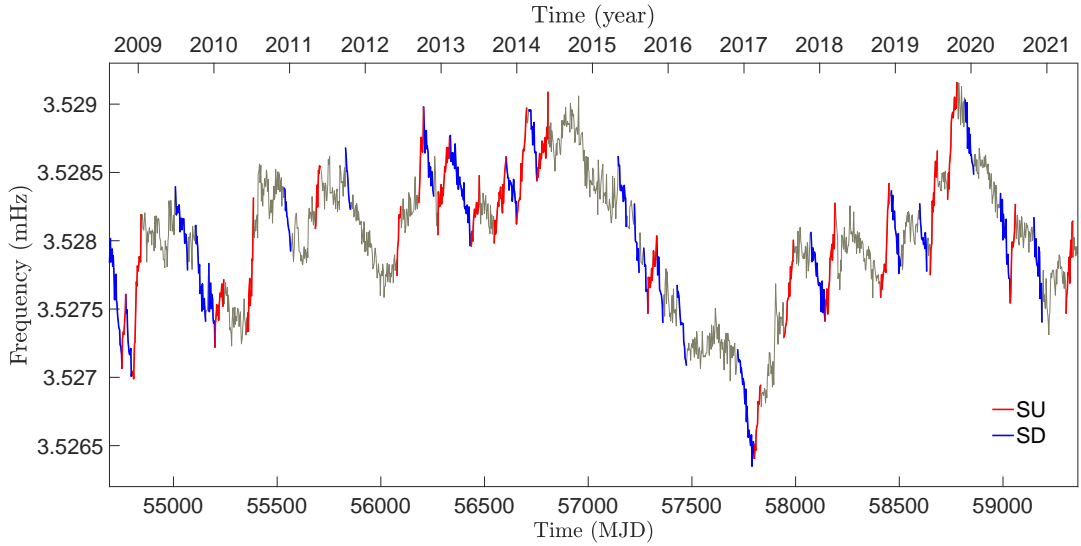}
    \caption{Decade-long spin frequency history of Vela X-1 monitored 
	 by \fer/GBM, with the selected SU (red) and SD (blue) intervals for following analysis.}
    \label{fig:mki}
\end{figure*}
%-------------------------------------------------------------
%-------------------------------------------------------------
\begin{figure}
    \centering
    \includegraphics[width=0.9\columnwidth]{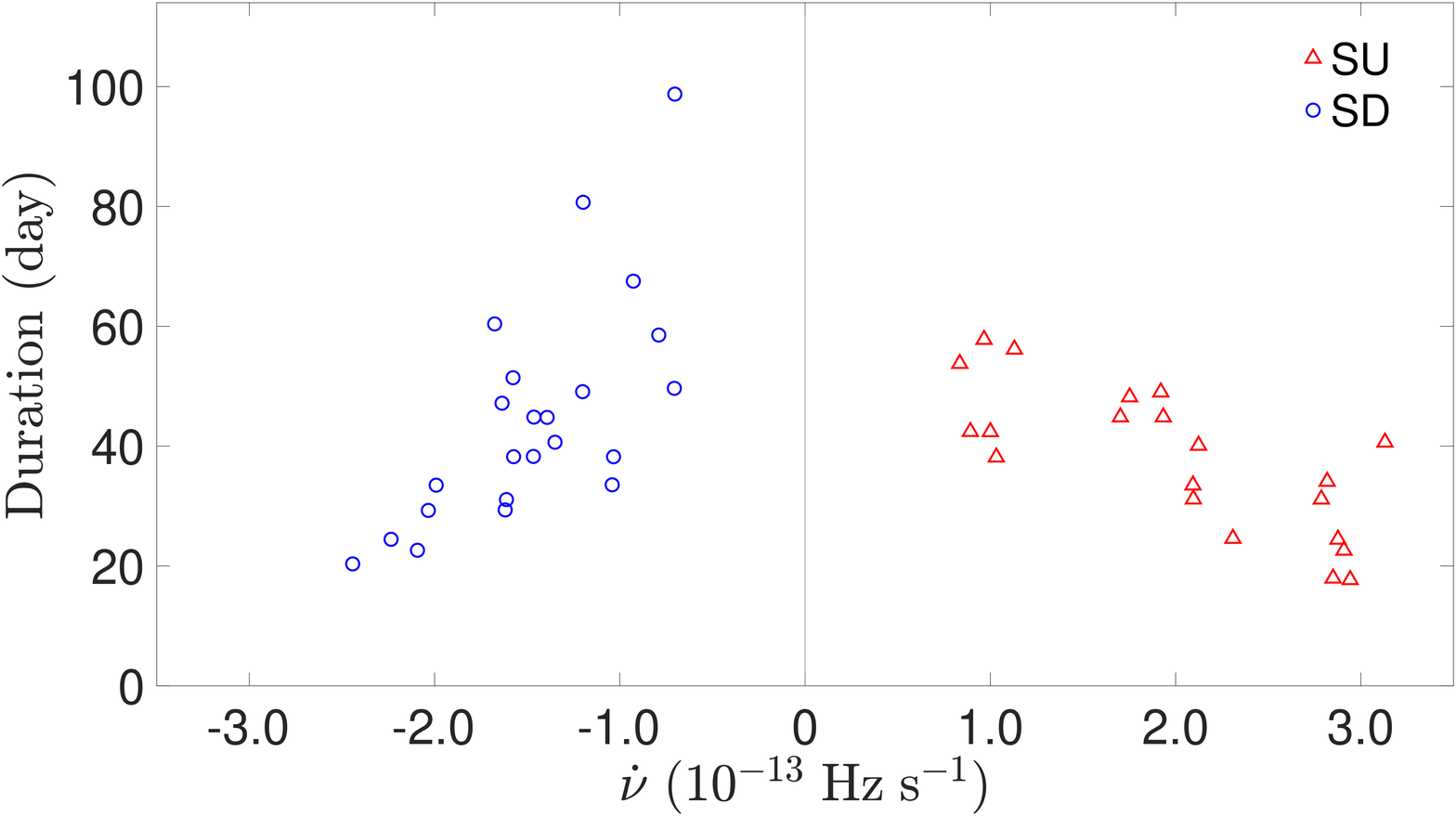}
    \includegraphics[width=0.9\columnwidth]{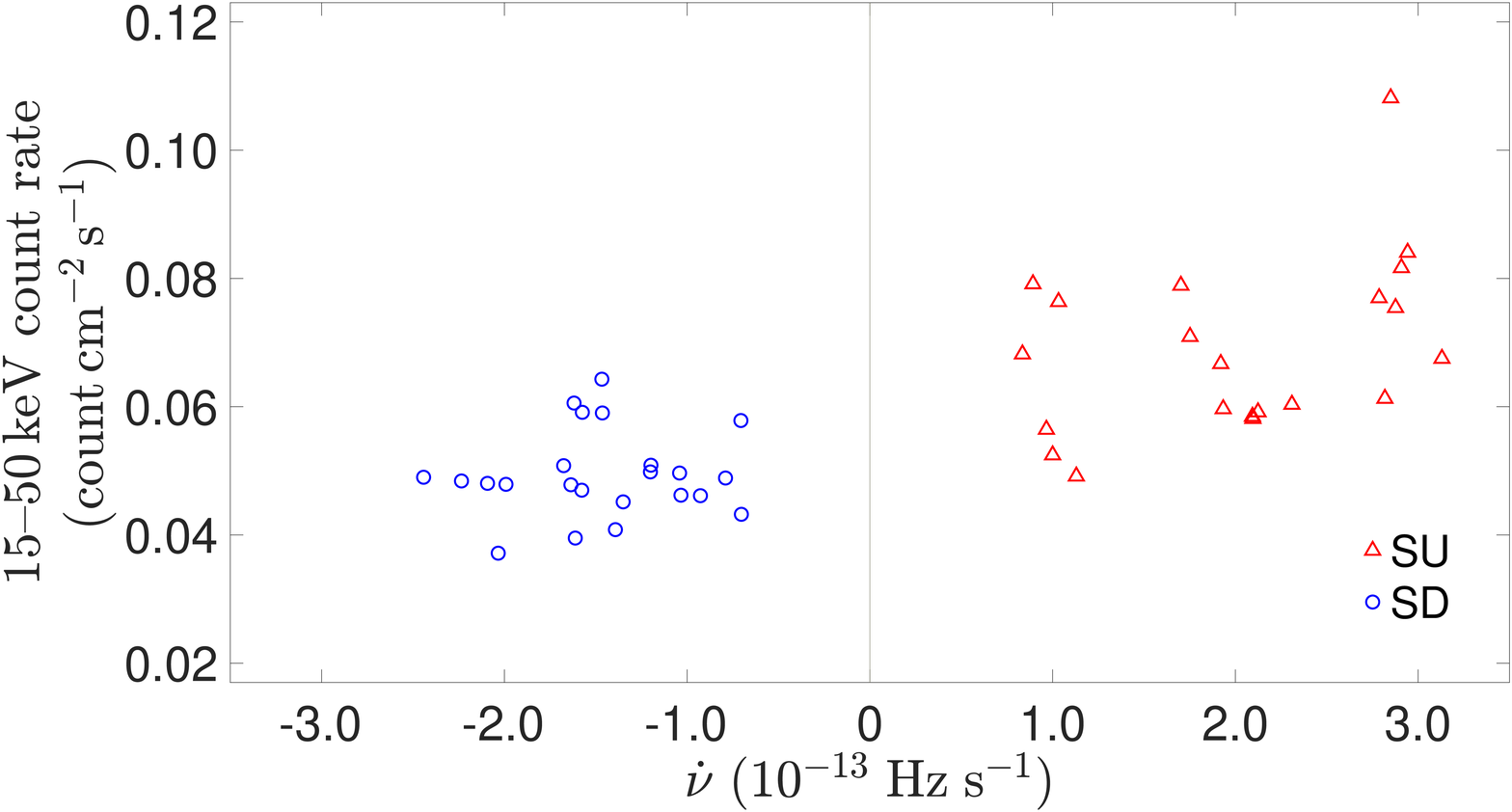}
	 \caption{The duration (top) and BAT flux (bottom) versus the 
	spin changing rate for SU/SD intervals.} 
    \label{fig:dmr}
\end{figure}

X-ray pulsars are magnetized neutron stars accreting material from their companion star
and producing regular X-ray pulses due to the spin of the neutron star.
Unlike radio pulsars that generally spinning down, the spin evolution of X-ray pulsars 
is more complicated. Many classical sources,
such as Vela X-1, OAO 1657-415, GX 301-2, Cen X-3, and GX 1+4, 
show alternating spin-up and spin-down
episodes on various timescale from days to decades, a phenomena not fully
understood yet
\citep[e.g.][]{Bild1997,Mal2020}.
In principle, the spin evolution of X-ray pulsars is governed by 
the interaction of the accreting flow with the magnetosphere and the neutron star.
The complex nature of this multi-scale, multi-physics interaction hampered our 
understanding of the spin behavior of X-ray pulsars,
especially for wind-fed systems \citep[e.g.][]{Mar17}.

In a recent study of OAO 1657-415, 
we found its spin frequency derivative
is correlated with luminosity during spin-up periods and anti-correlated with 
luminosity during spin-down periods \citep{Liao2022}, similar to 
GX 1+4 \citep{Chak1997}.
Moreover, we found that the orbital profile of OAO 1657-415 is dependent on the 
spin-up/spin-down state, indicating the accretion torque is related, somehow, with 
the property of accretion flow on the binary orbital scale.
This finding motivates us to check whether the orbital behavior of other 
X-ray pulsars is dependent on the spin-up/spin-down state of the neutron star.

In this letter, we perform a torque-dependent study of the orbital 
property of Vela X-1, an archetype of wind-fed X-ray pulsar. 
It has a spin period of $\sim283$\,s \citep{McC1976},
an orbital period of 8.9644\,days \citep{Fala2015, Krey2008},
a tight orbital separation of $53.4\,\rm R_{\sun}$ \citep{vanKer1995}, 
a low eccentricity $e \sim 0.09$ \citep[e.g.][]{Boy1986},
and a moderate magnetic 
field around $2.6\times10^{12}\,\rm G$ \citep[e.g.][]{Kret1996}.
Its optical companion, HD 77581, is an early-type B supergiant.
Many properties of Vela X-1 system was recently reviewed by \citet{Kret2021}.

\section{Observations}
\label{sec:instru}
On-board \fer~Gamma-ray Space Telescope, Gamma-ray Burst Monitor (GBM) is an 
unfocused and all-sky instrument, well suitable for monitoring
X-ray pulsars \citep{Meegan2009}.
The GBM Accreting Pulsar Program (GAPP) monitors dozens of accreting pulsars 
and publishes frequency and pulsed flux histories on its 
webpage\footnote{https://gammaray.msfc.nasa.gov/gbm/science/pulsars} \citep{Mal2020}.
To measure the spin frequency of Vela X-1, the eclipsing time was excluded, 
and the non-eclipsing time 
within one orbital cycle was divided into 3 segments. 
The spin frequency is searched in each segment based on $Y_{\rm n}$ statistic, 
and the frequency with $Y_{\rm n}$ above a certain value is considered a 
significant detection \citep[][]{Fin09}.

The Burst Alert Telescope (BAT) onboard {\it Neil Gehrels Swift Observatory} 
is designed to provide critical GRB triggers with a large field of view.
It has been continuously monitoring
the X-ray sky in 15--200\,keV band since 2005 \citep{Krimm2013}. 
The Monitor of All-sky X-ray Image (\maxi) is an high energy astrophysical 
experiment deployed on International Space Station in 2009. 
Its Gas Slit Camera \citep[GSC,][]{GSC} has been continuously monitoring 
the whole sky in 2-20\,keV band.
To study the long-term spin behavior of Vela X-1, we analyze the spin 
history of Vela X-1 monitored by GAPP, together with the corresponding light 
curves from \swi/BAT and \maxi/GSC.

\section{Results}
\label{sect:res}

%-------------------------------------------------------------
\begin{figure}
    \centering
    \includegraphics[width=0.5\textwidth]{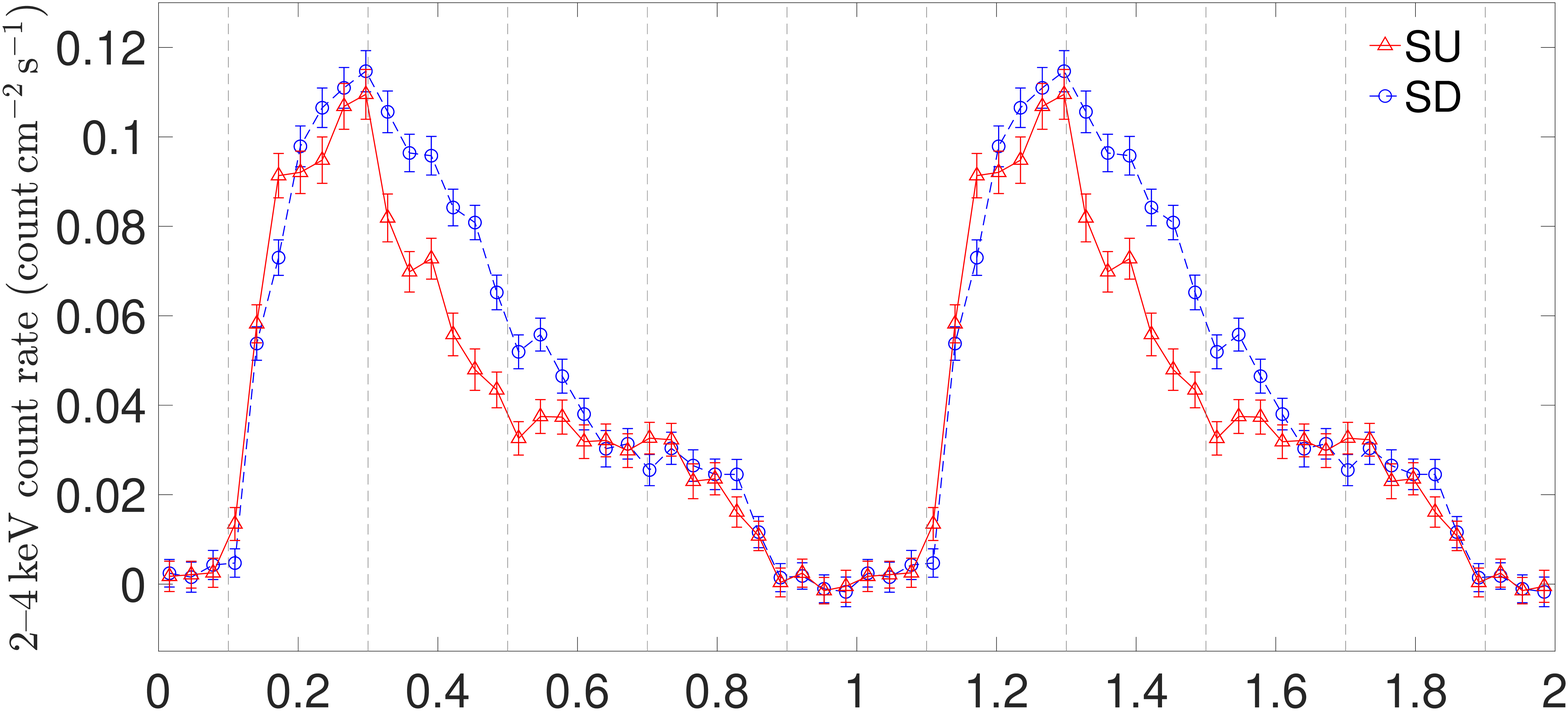}
    \includegraphics[width=0.5\textwidth]{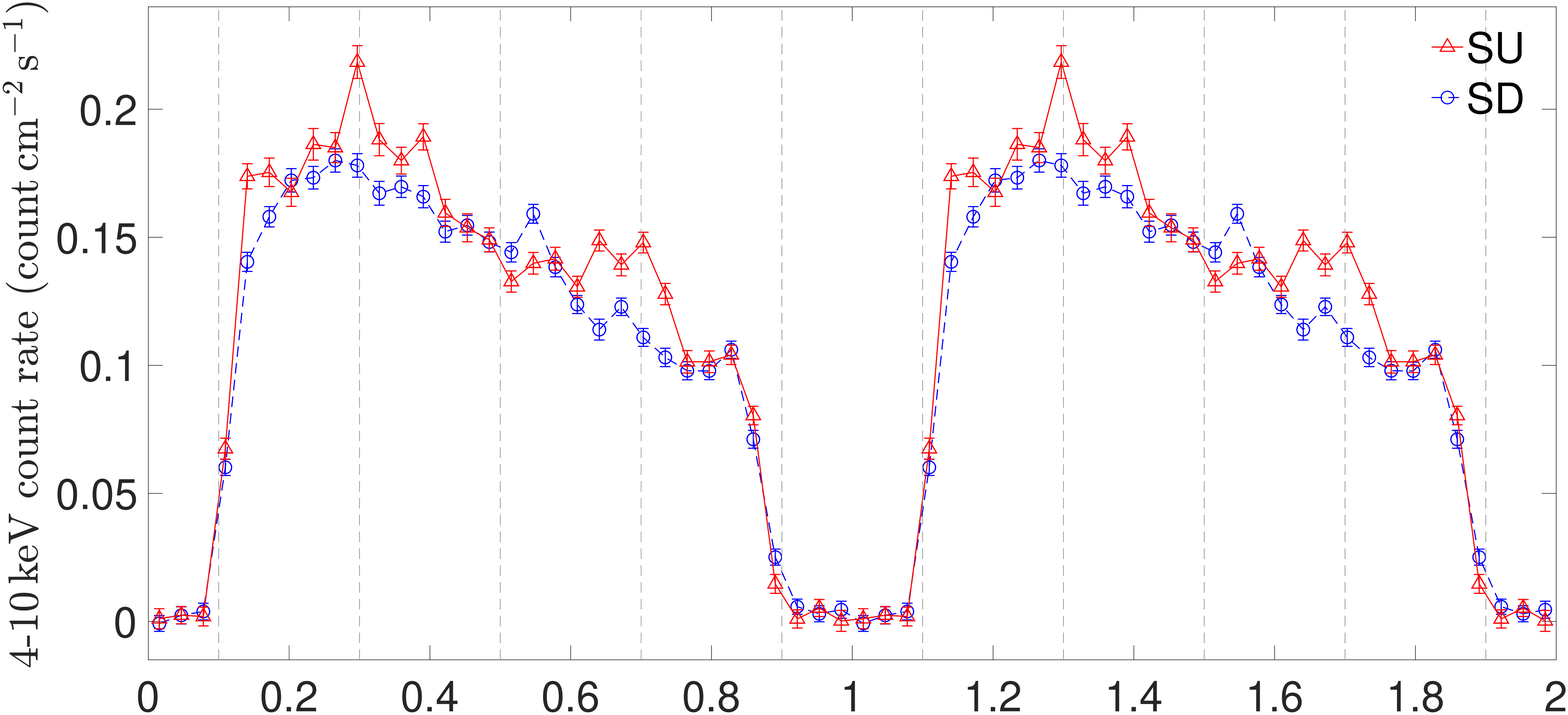}
    \includegraphics[width=0.5\textwidth]{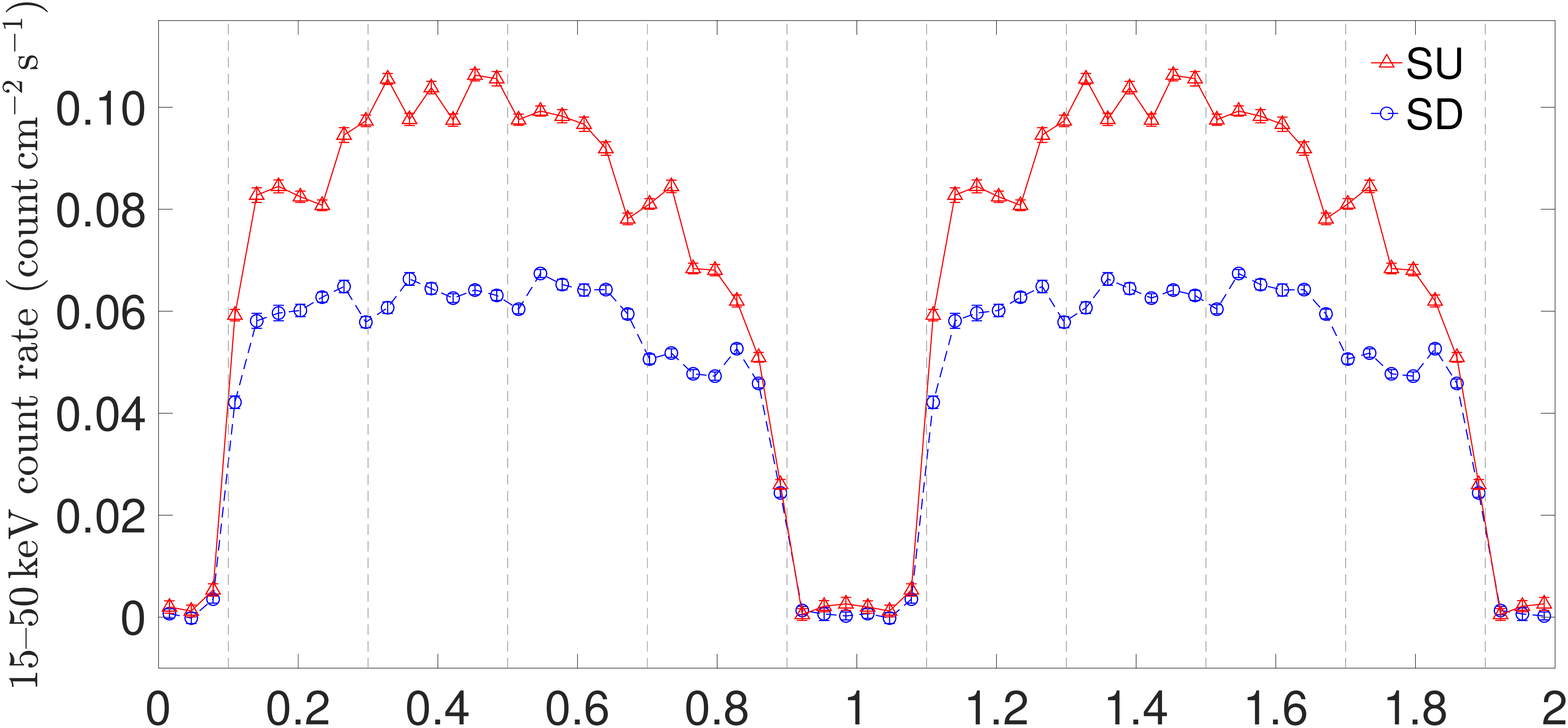}
    \includegraphics[width=0.5\textwidth]{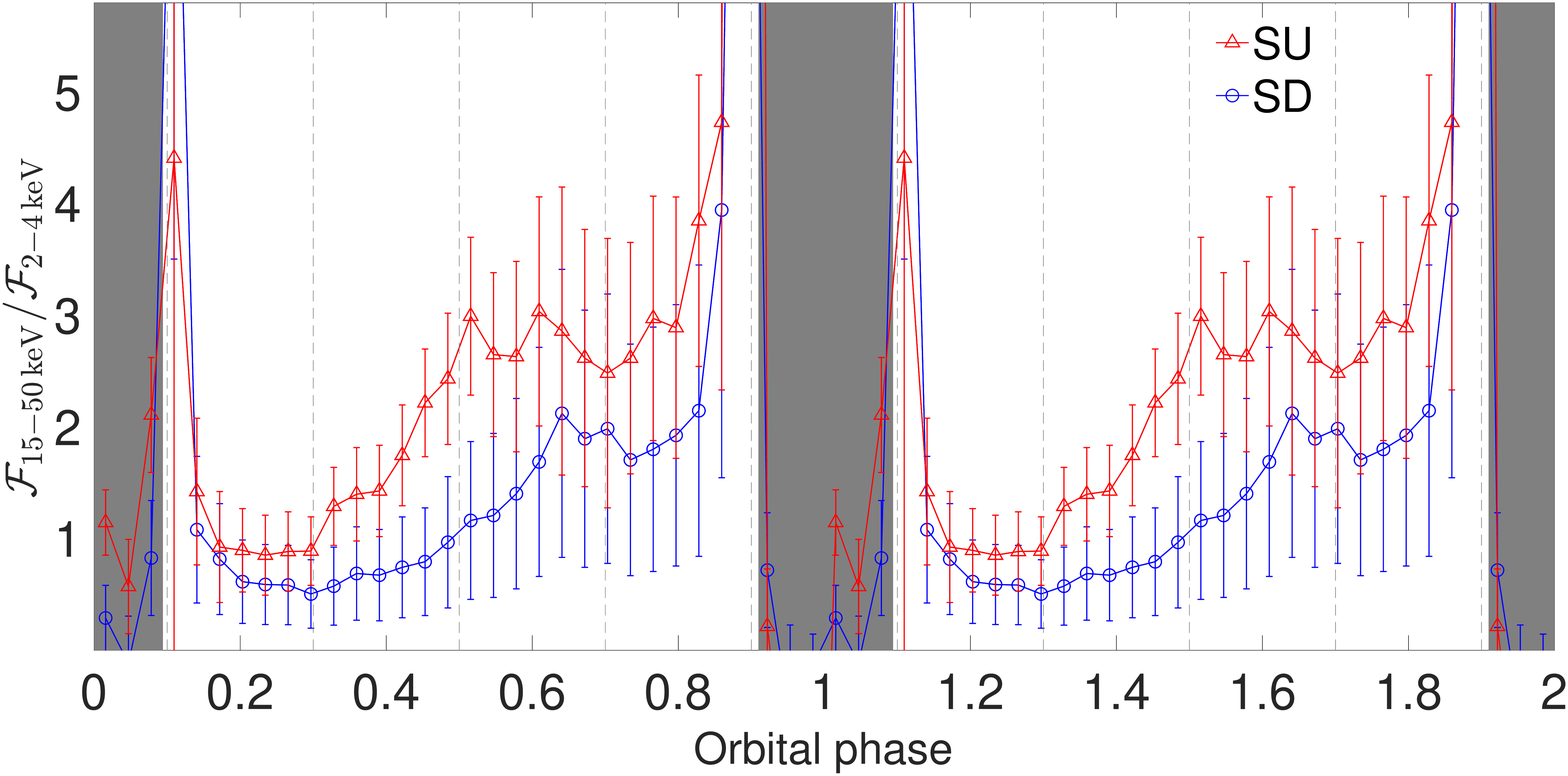}
    \caption{Orbital profile of Vela X-1 in soft band 
	 (2--4\,keV, top; 4--10\,keV, second) 
	 and hard energy band (15--50\,keV, third), and the hardness ratio (bottom)
	 for the selected SU and SD episodes. Vertical dashed lines mark the four phase bins 
	 used for spectral extraction.
    }
    \label{fig:opf}
\end{figure}
%-------------------------------------------------------------
The spin frequency of Vela X-1 monitored by \fer/GBM, 
together with the pulsed flux, is presented in Figure~\ref{fig:mki}.
As can be seen, the spin frequency varies on the shortest timescale of measurement.
The erratic spin variations of Vela X-1 had been found to be consistent 
with a random walk in pulse frequency \citep{Dee1989}. 
On the other hand, its spin history also 
shows some continuous spin-up/spin-down trend over tens of days. 
These continuous spin-up/spin-down periods have been pointed out 
by \citet{Mal2020} and \citet{Kret2021}
and been discussed roughly in a general sense.
Here we perform a detailed analysis of these periods. 
To analyze the spin-up (SU) and spin-down (SD) episodes separately, 
we first select, by eye, all apparently continuous SU/SD intervals 
longer than 2 orbital period (17.93 days). 
Then we find that most of the selected intervals have
a change of spin frequency $\Delta\nu>4.5 \times 10^{-7}\,\rm Hz$ and
an absolute value of Pearson coefficient larger than 0.8. So we remove 
a few intervals not satisfying these two conditions.
The selected SU/SD intervals are marked with red/blue color in Figure~\ref{fig:mki}.

We calculate an average spin frequency derivative within each interval via a linear fit.
The obtained rate is plotted against the duration of each interval
in the left panel of Figure~\ref{fig:dmr}. As can be seen, the periods of SU intervals 
last for 20--60 days, while those of SD intervals can last for a little longer. 
The spin changing rates in the SD intervals are generally
larger than $-2\times10^{-13}$\,Hz\,s$^{-1}$, while the largest spin changing rate
in the SU intervals reaches $3\times10^{-13}$\,Hz\,s$^{-1}$. 
The absolute value of the spin changing rate seems anti-correlated with the 
duration of each interval, for both SU and SD intervals (with a Pearson coefficient 
of -0.75 and 0.69, respectively).
In the right panel of Figure~\ref{fig:dmr}, we plot the average \swi/BAT flux in 15--50\,keV
against the spin changing rate, for the SU and SD intervals, respectively.
The average BAT flux in the SU episodes is about $0.07\,\rm count\,cm^{-2}\,s^{-1}$,
a little higher than that in the SD episodes ($0.05\,\rm count\,cm^{-2}\,s^{-1}$).
It seems the BAT fluxes and spin frequency derivatives are
too scattered to show a definite correlation, and 
the Pearson coefficient for SU/SD intervals is 0.39 and 0.09, respectively. 
%We also check the relation of spin frequency derivative with 
%BAT flux on the shortest timescale of measurement for the selected
%SU/SD intervals and find no apparent correlation.

\subsection{Orbital profile}

The orbital profile of Vela X-1 had been extensively studied in the literature, 
and it was found to be energy-dependent:
the higher the energy band, the simpler the shape \citep[e.g.][]{Kret2021}.
Based on the ephemeris in \citet{Fala2015}, we calculate the orbital profile
of Vela X-1 for the selected SU and SD episodes separately. 
The soft band (2--4\,keV, 4--10\,keV) and the hard band (15--50\,keV) 
profile is extracted from \maxi/GSC and \swi/BAT, respectively.
The obtained profiles are presented in Figure~\ref{fig:opf}.
Note that these profiles are averaged over multiple stretches of data, the 
individual profile of which could be quite different.

The soft band orbital profiles in 2--4\,keV are peaked around 
phase 0.3, similar as those reported in the literature. 
The 2--4\,keV profile of the SD intervals is a little higher 
(by a factor $\sim1.2-1.3$)
than that of the SU intervals around the orbital phases of 0.2--0.6, and they look 
similar at other phases.
The 4--10\,keV profiles are also peaked around phase 0.3, but 
are quite similar for both the SU and SD intervals.
The hard band profiles are more flat, which is also consistent with previous studies.
In contrast to the soft profiles, the hard fluxes in the SD episodes are lower 
than those in the SU episodes, by a factor $\sim0.6-0.7$, during the orbital phases
within 0.15--0.85. 

From the orbital profiles within the soft and hard bands, it can be seen that 
there are spectral differences between the SU and SD episodes. 
The orbital hardness ratio profile between the hard and soft bands is shown in 
the bottom panel of Figure~\ref{fig:opf}. The hardness ratio in the SU episodes
is larger than 
that in the SD episodes for phases out of eclipse. 
Around the eclipse phases (shaded areas), the hardness ratio shows large 
fluctuation due to the very small fluxes.

\subsection{Spectra}
%-------------------------------------------------------------
\begin{figure*}
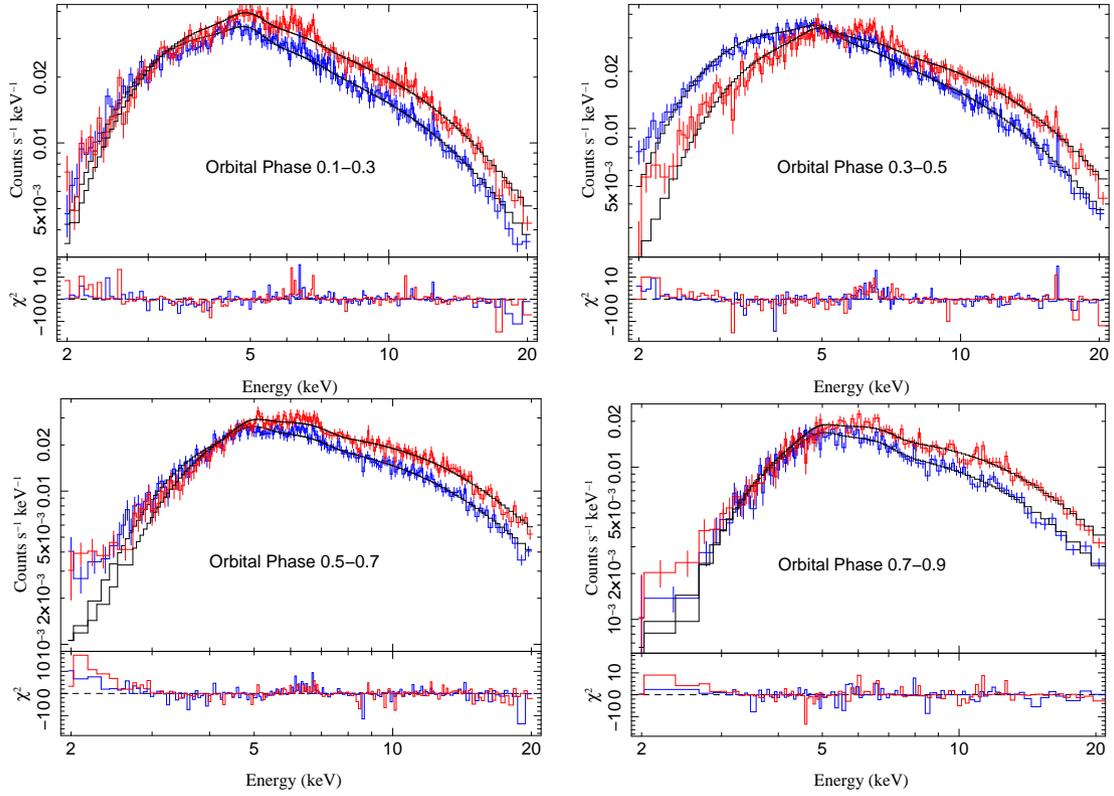

    \includegraphics[width=0.6\columnwidth, angle=270]{fig/cud_phi1.eps}
    \includegraphics[width=0.6\columnwidth, angle=270]{fig/cud_phi2.eps}
    \includegraphics[width=0.6\columnwidth, angle=270]{fig/cud_phi3.eps}
    \includegraphics[width=0.6\columnwidth, angle=270]{fig/cud_phi4.eps}
	 \caption{Extracted spectrum in each orbital phase bin for
	 the SU (red) and SD (blue) 
	episodes, together with the fitted model (black). }
    \label{fig:spec}
\end{figure*}
%-------------------------------------------------------------
%-------------------------------------------------------------
\begin{figure*}
    \centering
    \includegraphics[width=1\columnwidth]{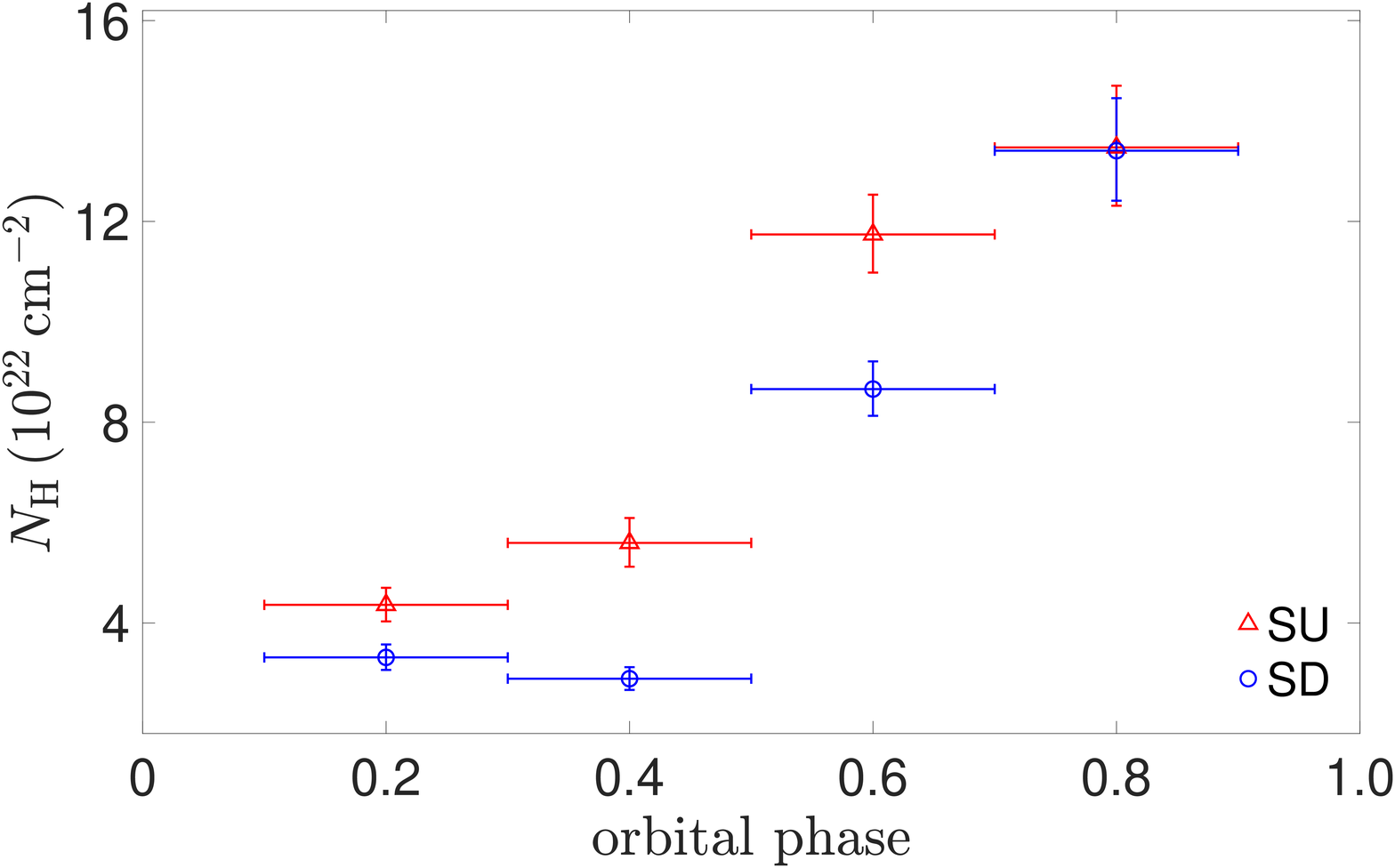}
    \includegraphics[width=1\columnwidth]{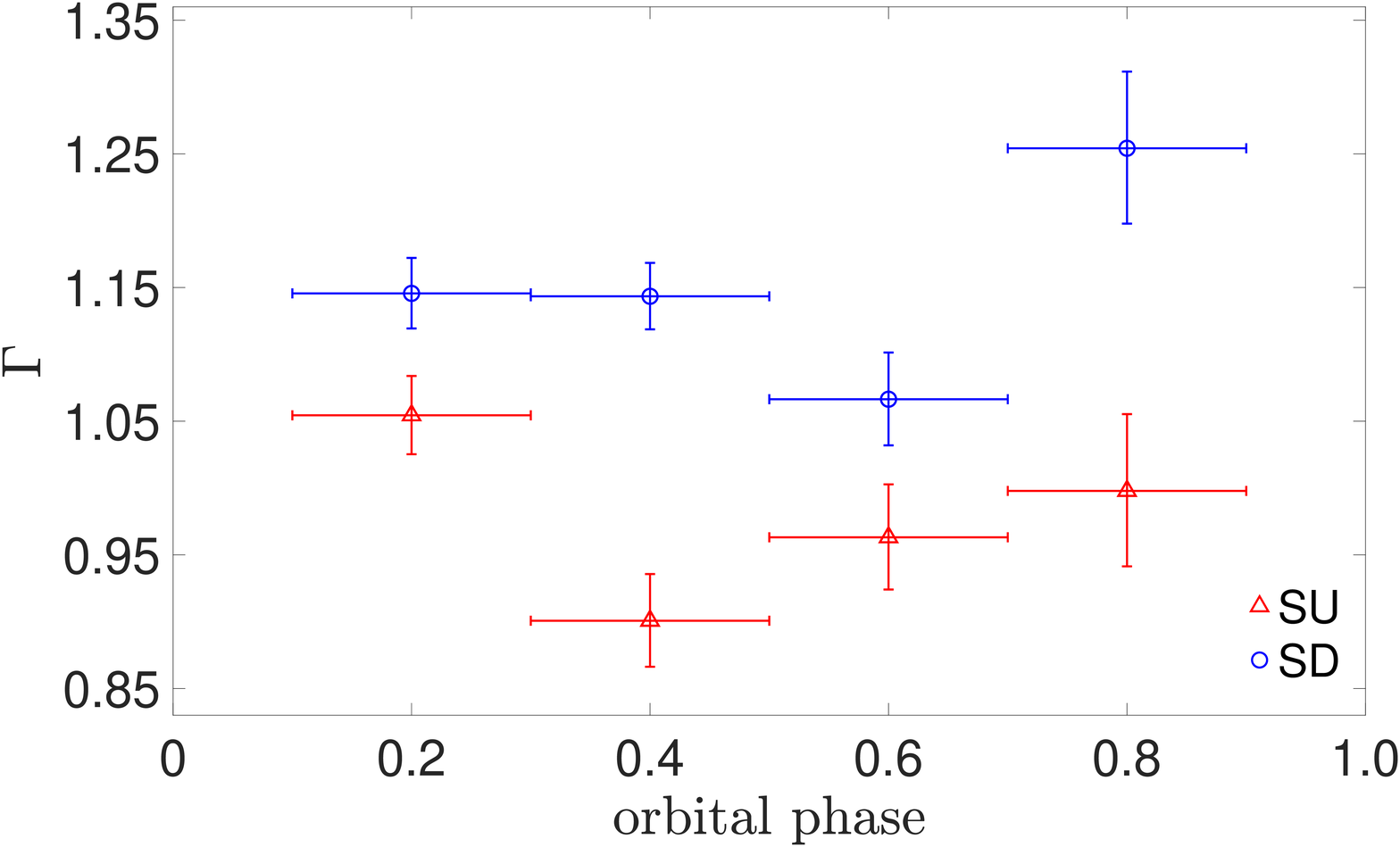}
	 \caption{The fitted column density (left) and photon index (right) 
	in each orbital phase bin for the SU (red) and SD (blue) episodes.}
    \label{fig:nhpi}
\end{figure*}
%-------------------------------------------------------------
%-------------------------------------------------------------
\begin{table*}
    \renewcommand\arraystretch{1.2}
    \centering
	\caption{Fitting results of an absorbed power-law ($\mathrm{phabs\times powerlaw}$). Errors are given in 90\% confidence level.}
    \begin{threeparttable}
    \begin{tabular}{cccccc}
    \hline
    $\phi_{\rm orb}$ & $N_{\rm H}\,(10^{22}\,\rm cm^{-2}$) & $\Gamma$ & Norm\tnote{1} & Luminosity\tnote{2}~~$(10^{36}\,\rm erg/s)$ & {$\chi^2_\nu$} \\
    \hline
    SD 0.1--0.3 & $3.32\pm0.25$ & $1.15\pm0.03$ & $0.24\pm0.01$ & $5.87\pm0.44$ & 1.32 \\
    SD 0.3--0.5 & $2.89\pm0.23$ & $1.14\pm0.02$ & $0.23\pm0.01$ & $5.95\pm0.25$ & 1.62 \\
    SD 0.5--0.7 & $8.66\pm0.54$ & $1.07\pm0.03$ & $0.21\pm0.02$ & $5.88\pm0.24$ & 1.48 \\
    SD 0.7--0.9 & $13.4\pm1.0$ & $1.25\pm0.06$ & $0.20\pm0.03$ & $4.25\pm0.19$ & 1.28 \\
    SU 0.1--0.3 & $4.36\pm0.34$ & $1.05\pm0.03$ & $0.24\pm0.02$ & $7.42\pm0.31$ & 1.53 \\
    SU 0.3--0.5 & $5.60\pm0.49$ & $0.90\pm0.03$ & $0.17\pm0.02$ & $7.15\pm0.21$ & 1.44 \\
    SU 0.5--0.7 & $11.7\pm0.80$ & $0.96\pm0.04$ & $0.20\pm0.02$ & $7.62\pm0.30$ & 1.66 \\
    SU 0.7--0.9 & $13.5\pm1.2$ & $1.00\pm0.06$ & $0.16\pm0.03$ & $5.17\pm0.27$ & 1.24 \\
    \hline
    \end{tabular}
    \begin{tablenotes}
        \footnotesize
        \item[1] Normalization of the power-law, in units of $\rm photon\,keV^{-1}\,cm^{-2}\,s^{-1}$ at 1\,keV.
        \item[2] Absorption-corrected luminosity within 2--30\,keV band.
    \end{tablenotes}
    \end{threeparttable}
    \label{tab:para}
\end{table*}
%-------------------------------------------------------------

To further illustrate the spectral differences between different torque states, 
we divide the orbital phase into four bins (0.1--0.3, 0.3--0.5, 0.5--0.7, 0.7--0.9), 
and extract the spectrum of each phase bin from the accumulated event files 
provided by \maxi/GSC, for SU and SD intervals, respectively.
The orbital-phase-resolved spectra are plotted in Figure~\ref{fig:spec}. 
Note that these spectra are averaged over data of different periods and flux 
levels, and the individual spectrum could be quite different.

As can be seen, all the spectra of the SU intervals 
are higher than those of the SD intervals above 5 keV. While below 5 keV, 
the spectra of the SU intervals are apparently lower than those of the SD intervals
for phases within 0.3-0.5, consistent with 
the results of the orbital profiles.
To quantify the spectral differences, we 
fit a simple model of absorbed power-law ($\mathrm{phabs\times powerlaw}$)
to the spectra. The fitted spectra are plotted as black solid histograms 
in Figure~\ref{fig:spec}, and the fitting parameters are listed in Table~\ref{tab:para}.
For the viewing purpose, the fitted column densities and photon indexes 
are also plotted in Figure~\ref{fig:nhpi}.

The obtained column densities of the SU intervals are generally higher than 
those of the SD intervals, which is most apparent for orbital phases between 0.3 and 0.7.
This is consistent with 
the relatively higher hard flux and lower soft flux of the SU intervals. 
On the other hand, the photon indexes of the SU intervals are generally below
those of the SD intervals.
We note that below 3 keV, the model fluxes are lower than the observed 
fluxes, especially for the later orbital phases between 0.5 and 0.9.
This may indicate inhomogeneous absorption and/or emergence of emission lines.
The fitted column densities are a little lower than those obtained by 
individual spectral studies \citep[e.g. Fig. 5 in][]{Kret2021}, which may be 
due to the simplified modelling of the averaged spectra.
Adopting a distance of $2\rm\,kpc$ derived from the third {\it Gaia} data 
release \citep{Kret2021}, the absorption-corrected luminosity within 2--30\,keV 
band is estimated in each orbital phase bin, and the results are listed
in Table~\ref{tab:para}. The intrinsic luminosity of the SU intervals is 
generally higher (with a factor $\sim10-20\%$) than that of the SD intervals.

\section{Discussions and Conclusions}
\label{sect:dc}
Based on the long-term data monitored by \fer/GBM, \swi/BAT and \maxi/GSC, we studied 
the orbital profile of Vela X-1, for the 
intervals of different torque states (SU/SD).
The orbital profile of the SU intervals in 15--50 keV band is systematically higher 
than that of the SD intervals, while the profile of the SU intervals in 2--4 keV
is lower than that of the SD intervals around phase 0.2--0.6.
A detailed spectral analysis showed that the fitted column densities during the 
SU intervals are generally higher than those of the SD intervals, while 
the fitted photon indexes of the SU intervals are generally smaller 
than those of the SD intervals.

The torque-dependent column density came out as a surprise. 
The periods of selected SU/SD intervals are around 20--100 days, which 
are longer than the orbital period of Vela X-1 (8.9644 days). 
The higher column density around orbital phase 0.6--0.9 could be explained 
as due to a trailing stream \citep[e.g.][]{Dor13}, which comes into the line of 
sight around phase 0.5-0.6 with probable variations in individual orbit and 
may explain the quite different column densities found at this phase range.
The fact, that the column densities of the SU intervals 
around both phases of 0.3 and 0.6 are higher than those of the SD intervals,
indicates that there is more material between the neutron star and 
the earth during the SU intervals than the SD intervals, 
regardless of the position of the stream.

Then, the problem is on what spatial scale the enhanced/reduced material during 
the SU/SD intervals occurred? The involved time scale of the SU/SD intervals
of 20--100 days is much longer than the dynamical time scale on the gravitational 
capture (Bondi) radius, also longer than the cooling time scale as involved 
in the quasi-spherical accretion model \citep{Shaku2012}, or 
the flip-flop timescale \citep[e.g.][]{Mat1987,Blon1990}.
Therefore, it seems the most feasible scale of the enhanced/reduced material during 
the SU/SD intervals is the orbital scale. That is, the wind material 
itself is likely enhanced/reduced during the SU/SD intervals.

The stellar wind of massive stars is subjected to line-driven-instability
\citep{OR84} and is composed of small clumps
\citep[for a recent review, see][]{Mar17}. Apparently, the spatial scale 
of the enhanced/reduced material during the SU/SD intervals is
much larger than those of individual clumps. 
Recently, \cite{Chan2021} 
reported a possible periodic variation of spin period of Vela X-1 over a 
time $\sim$5.9\,years,
which may be due to cyclic activity of the supergiant companion, HD 77581.
The different wind material during the SU/SD intervals of Vela X-1 
could also be originated from HD 77581, if its wind is varying
on tens of days and part of the accreting matter has a specific angular 
momentum similar as that of orbital matter \citep{IS75}. The sign of the 
angular momentum of the accreting matter depends on whether the matter is accreted 
prograde or retrograde to the neutron star, which produces the 
observed SU/SD trend.

The recent estimations of the wind speed at the distance of Vela X-1 by 
\citet{Gime2016} and \citet{Sand2018} were
close to or even lower than the orbital velocity of Vela X-1 
\citep[see Fig. 19 in][]{Kret2021}. The small wind speed will make the wind
heavily affected by the orbiting neutron star and make the formation
of a transient wind-fed disk possible \citep{Kari2019,Mell2019,Liao2020}.
While the BAT flux of the SD intervals, on average, is a little lower than that 
of the SU intervals, the relation between BAT flux and spin frequency derivative is 
too scattered to show a significant correlation. This implies that 
the luminosity is dominated by accreted matter of negligible angular momentum, 
not that transferring angular momentum to the neutron star.
This is a case quite different from OAO 1657-415, for which correlation/anti-correlation
between the flux and spin frequency derivative is observed for SU/SD intervals.
For accretion through a standard thin disk, the predicted spin-up 
rate of Vela X-1 is around $8\times10^{-13}$\,s$^{-2}$ for a luminosity 
$\sim6\times10^{36}$erg\,s$^{-1}$ \citep[e.g.][]{Liao2020}, 
about 3-5 times larger than the average value of the selected intervals.
This indicates that the accretion flow of Vela X-1
should be more irregular/spherical compared 
with that of OAO 1657-415, as also evidenced by the more erratic spin variations 
of Vela X-1 than OAO 1657-415. For spin-up/spin-down rates of other 
accretion scenarios (such as quasi-spherical model), we refer to 
\citet[][]{Kret2021} for a detailed discussion.

The fitted photon indexes of the SU intervals are systematically lower than 
those of the SD intervals. This might be related with the interaction between 
the prograde/retrograde flow with the magnetic field of the neutron star, 
and further investigation of their connection is needed to test it.

In summary, the SU intervals of Vela X-1 show a larger luminosity/accretion rate 
and higher column density than the SD intervals. It could be caused by the 
variations of the stellar wind of HD 77581 on tens of days, which lead to 
alternating prograde/retrograde accreting flow to the neutron star. Th
prograde/retrograde flow dominates the transfer of the angular momentum to 
Vela X-1, but not the total observed luminosity. 
If this scenario is true, one may expect other observable signature of the 
varying stellar wind, such as the varying P-Cygni profile of UV resonance lines,
in different torque states. The result indicates that the stellar wind
of massive stars, besides a clumpy structure, 
could be variable on tens of days.

\section*{Acknowledgements}
We thank our referee for comments improving much of the manuscript.
JL acknowledges the support by National Science Foundation of China (NSFC U1938113) and by the Scholar Program of Beijing Academy of Science and Technology (DZ BS202002).
%LG acknowledges the support by the National Program on Key Research and Development Project (2016YFA0400804) and by the Strategic Priority Research Program of the Chinese Academy of Sciences (XDB23040100).

%%%%%%%%%%%%%%%%%%%%%%%%%%%%%%%%%%%%%%%%%%%%%%%%%%
\section*{Data Availability}
The data utilized in this article are observed by {\it Fermi}/GBM, {\it Swift}/BAT and {\it MAXI}/GSC and are publicly available.
%\newline
%\href{https://gammaray.nsstc.nasa.gov/gbm/science/pulsars.html}{https://gammaray.nsstc.nasa.gov/gbm/science/pulsars.html}, 
%\href{https://swift.gsfc.nasa.gov/results/transients/VelaX-1}{https://swift.gsfc.nasa.gov/results/transients/VelaX-1}, \href{https://maxi.riken.jp/star\_data/J0902-405/J0902-405.html}{https://maxi.riken.jp/star\_data/J0902-405/J0902-405.html}.

\bibliographystyle{mnras}
\bibliography{vx1ud} % if your bibtex file is called example.bib

%\begin{thebibliography}{99}
%\end{thebibliography}

%%%%%%%%%%%%%%%%%%%%%%%%%%%%%%%%%%%%%%%%%%%%%%%%%%

%%%%%%%%%%%%%%%%% APPENDICES %%%%%%%%%%%%%%%%%%%%%

\appendix

% \section{Some extra material}
% If you want to present additional material which would interrupt the flow of the main paper,
% it can be placed in an Appendix which appears after the list of references.

%%%%%%%%%%%%%%%%%%%%%%%%%%%%%%%%%%%%%%%%%%%%%%%%%%

% Don't change these lines
%\bsp	% typesetting comment
\label{lastpage}
\end{document}